\documentclass[fleqn,10pt]{wlscirep}
\usepackage[utf8]{inputenc}
\usepackage[T1]{fontenc}
\usepackage[sort&compress,numbers]{natbib}
\usepackage{listings}
\usepackage{cleveref}
\usepackage{graphicx}
\usepackage{subcaption}
\usepackage{xr}
\DeclareUnicodeCharacter{202F}{\,}   
\newcommand{\tref}[1]{Table~S\ref{#1}}
\usepackage{longtable}

\title{Scientific mobility patterns of Indian researchers: Impact on career growth}

\author[1]{Siraj TM}
\author[2]{Harikrishnan S}
\author[3]{Mathew Vincent}
\author[4]{Sandeep Chowdhary}
\author[5]{Chandrakala Meena}

\affil[1]{School of Data Science, Indian Institute of Science Education and Research Thiruvananthapuram, 695551, India}
\affil[2]{School of Mathematics, Indian Institute of Science Education and Research Thiruvananthapuram, 695551, India}
\affil[3]{School of Physics, Indian Institute of Science Education and Research Thiruvananthapuram, 695551, India}
\affil[4]{International Institute of Applied Systems Analysis, A-2361, Laxenburg, Austria}
\affil[5]{Department of Physics,  Indian Institute of Science Education and Research  Pune, 411008, India}

\begin{abstract}

Scientific mobility plays a pivotal role in shaping individual research careers and enhancing national innovation capacity by enabling knowledge exchange, fostering international collaborations, and providing access to leading research environments. Studying international mobility patterns of researchers from developing countries offers vital insights into strengthening domestic scientific ecosystems, addressing talent migration, and promoting global integration. In this study, we analyze the international mobility of India-affiliated researchers by reconstructing longitudinal affiliation trajectories from the OpenAlex database. We assembled longitudinal affiliation histories for 157,471 India-affiliated researchers and categorized them into distinct mobility pathways: immobile, returnees, and those who settled abroad after initially moving to the United States, European Union, or other high-income countries. Our analysis reveals that 28\% of Indian researchers who started their careers in India experience at least one international move, yet more than 73\% never return, highlighting a persistent pattern of brain drain. Notably, internationally mobile researchers predominantly originate from premier Indian institutions. These findings underscore the challenges of retaining top talent domestically and point to the urgent need for policies that incentivize return migration and strengthen domestic research infrastructure to leverage the expertise of globally trained scientists.
Matched pair analyses demonstrate that international mobility is associated with lasting benefits such as a sustained increase in citation impact, while publication rates generally realign with those of immobile peers. Mobility also substantially enhances international collaboration: the share of foreign co-authors increases from a 52\% baseline among immobile researchers to 83–87\% at the time of transition abroad, and remains elevated among returnees, who retain a long-term collaboration premium of 32–40 percentage points across research disciplines. Notably, returnees maintain global networks, serving as critical bridges connecting Indian science with global research systems. Further, we find that our results are robust across all major research disciplines, demonstrating consistent patterns of mobility, citation impact, and international collaboration regardless of the field of study.
Overall, our study underscores the dual role of scientific mobility as both a driver of scientific excellence and international engagement, and as an ongoing challenge for developing nations seeking to reintegrate returning talent.

\end{abstract}

\begin{document}
\flushbottom
\maketitle
\thispagestyle{empty}
\section*{Introduction}

Scientific mobility, the movement of researchers across institutional and national boundaries, plays a central role in shaping global knowledge production. It facilitates the exchange of ideas and skills, supports the formation of new research teams, and expands collaborative networks. As a result, mobility has a profound influence on both national research capacity and international scientific visibility\cite{franzoni2012foreign, appelt2015factors}. Mobility patterns are often driven by opportunity-seeking behavior, particularly the availability of research funding \cite{simini2012universal}, and are further shaped by cultural and linguistic affinities \cite{gargiulo2014mobility}. Although digital communication has lowered some barriers to collaboration, geographic distance remains a significant factor in structuring scientific cooperation and the diffusion of knowledge \cite{pan2012world}. For early-career researchers, mobility offers critical pathways to access elite institutions and global networks \cite{ma2020mentorship}, with early collaborations with prominent scientists serving as strong predictors of long-term research success \cite{li2019early}.

The concept of 'bridge institutions', those that connect disparate parts of the global scientific system, is particularly relevant in this context \cite{newman2004coauthorship}. Such institutions often serve as intermediaries, helping researchers from less connected regions integrate into the core of global science \cite{deville2014career}. Numerous studies have shown that international mobility enhances citation impact \cite{sugimoto2017scientists}, number of publications \cite{SongGan2022}, collaborative diversity \cite{wagner2015international}, and broadens exposure to novel research practices and perspectives \cite{robinson2019many}. Research has consistently found that internationally mobile scientists tend to achieve higher citation impact than their non-mobile peers \cite{franzoni2014mover, sugimoto2017scientists}, although this may partly reflect underlying selection effects in opportunity structures. The academic hiring system itself exhibits a hierarchical and geographically constrained structure, with most career transitions occurring early and within proximate locations \cite{clauset2015systematic, deville2014career}. At the macro level, mobility studies have frequently focused on quantifying brain drain and brain gain across countries and regions \cite{van2012global, arrieta2017quantifying}.

Several recent work have used large-scale data sources to study mobility patterns in various contexts, in particular, analyses of ORCID records from U.S. faculty careers have identified distinctive institutional transition patterns \cite{yan2020analyzing}, while studies of doctoral mobility suggest that even short-term international stays can produce durable changes in collaboration networks \cite{horta2021research}. Regional studies in the Middle East and North Africa \cite{el2021analyzing}, Italy \cite{abramo2022effect}, and China \cite{huang2024talent} underscore the impact of mobility on research performance and career progression. Despite this growing literature, scientific mobility in the context of emerging research systems such as India remains understudied. 

Previous investigations focusing on India\cite{Singh2019} have primarily relied on domain-specific or restricted datasets (e.g., American Physical Society publications), offering only partial insights into the full arc of researchers’ careers. Moreover, the relationship between international mobility, research impact, and collaborative networks remains underexplored in the context of the Indian scientific ecosystem.
To address these gaps, we analyze a comprehensive dataset comprising the career trajectories of $157,471$ researchers extracted from the OpenAlex corpus \cite{priem2022openalex}, focusing on authors with at least one India-affiliated publication. 
Based on longitudinal affiliation data, we categorize researchers into distinct mobility classes—non-mobile (those who never left India), those who settled abroad, and those who returned to India after a stay abroad. We focus only on researchers who moved to the United States (US), European Union (EU), or other high-income countries. This allow a systematic assessment of how different mobility pathways shape career outcomes. Our analysis addresses four research questions: \textit{What is the extent of brain drain from India; that is, what proportion of internationally mobile scientists return, and at what career stage does international movement typically occur? Is international mobility associated with a sustained increase in citation impact? Does mobility affect annual publication productivity? Do returnees maintain higher levels of international collaboration than non-mobile peers, thereby acting as conduits for global scientific engagement?}
We evaluate these questions using three career-age-normalized metrics: citations per paper per year, papers per year, and the proportion of co-authors affiliated outside India. To ensure robustness, we replicate our analyses across multiple disciplinary fields.

\section*{Methods}
\subsection*{Data Collection and Processing}
 To comprehensively analyze the dynamics of researchers who transition abroad and later return to India or settle abroad, it is essential to study a sufficiently large sample of data that provides meaningful insights into the Indian research community. Therefore, we obtain data from the OpenAlex repository of scholarly works, which offers structured metadata on publications, authors, institutional affiliations, and collaboration networks. From this dataset, we selected all authors who had been affiliated with at least one institution in India at any point during their research careers.
To filter out anomalous or unrepresentative cases, we restricted our analysis to researchers with a publication count between 10 and 200. This yielded a final dataset of $157,470$ authors who had been affiliated with Indian institutions at some stage of their professional journey.

For each of these authors, we sorted their publications in chronological order and used the associated institutional affiliations to reconstruct their career trajectories. Consecutive occurrences of the same institution across two publications were treated as evidence of continuous affiliation. To further refine the dataset, we excluded publications that were spaced more than three standard deviations above the mean publication gap for a given author, thereby removing outliers that may distort career path inference.

This approach enabled the construction of comprehensive institutional mobility timelines for each researcher, forming the foundation for our subsequent analyses of mobility categories, citation impact, productivity, and collaboration. An illustrative example is shown in Table \ref{table:Nissim_Career}, which presents the reconstructed career trajectory of Prof.~Dr.~Nissim Kanekar, an astrophysicist and cosmologist currently working at the Tata Institute of Fundamental Research (TIFR), India.

\vspace{0.25cm}
\begin{table}[hbt!]
\centering
\begin{tabular}{rllll}
\hline
 \textbf{\small{Moves}}   & \textbf{\small{Name}}                                & \textbf{\small{Country}}   & \textbf{\small{Start\ Date}}   & \textbf{\small{End Date}}   \\
\hline
  0 & National Centre for Radio Astrophysics & IN        & 1997-12-01   & 2002-02-01 \\
  1 & University of Groningen                & NL        & 2003-02-14   & 2005-01-01 \\
  2 & National Radio Astronomy Observatory   & US        & 2005-03-01   & 2018-05-16 \\
  3 & Tata Institute of Fundamental Research & IN        & 2018-06-07   & 2024-02-01 \\
\hline
\end{tabular}
\caption{The career trajectory of Prof. Dr. Nissim Kanekar, reconstructed using data from OpenAlex.}
\label{table:Nissim_Career}
\end{table}
\vspace{0.25cm}
In addition to tracking institutional affiliations, we analyzed co-author relationships to investigate collaboration dynamics. For each publication, we extracted detailed co-author data, including names, OpenAlex author IDs, affiliated institutions, country codes, and other relevant attributes and citation counts. 

\subsection*{Scientific Mobility Trajectories of Researchers}
\begin{figure}[h!]
    \centering

    \begin{subfigure}[b]{0.49\textwidth}
        \centering
        \includegraphics[width=\textwidth ,trim=0mm 1mm 0mm 0mm]{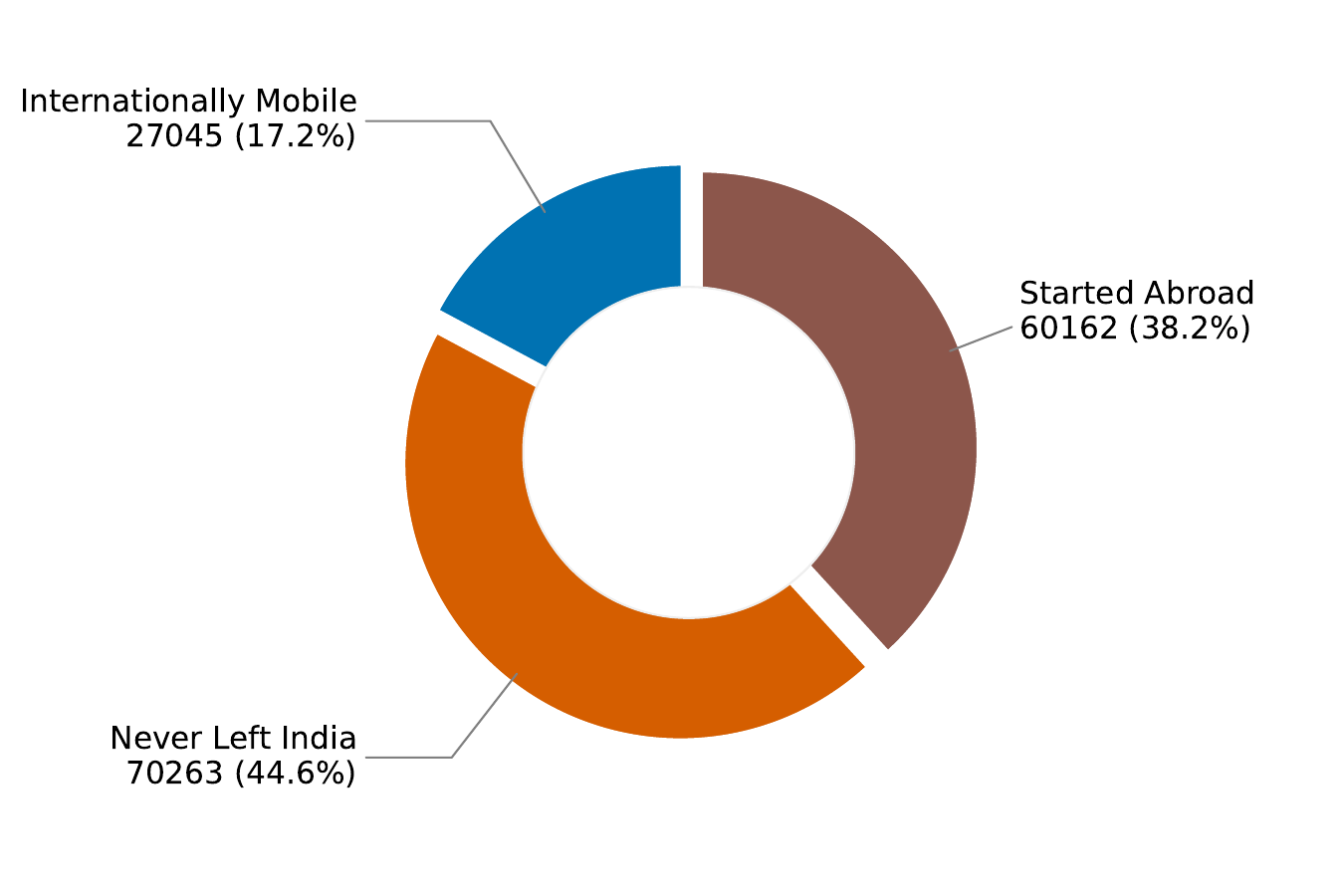}

        \label{fig:first_pie}
    \end{subfigure}
    \begin{subfigure}[b]{0.49\textwidth}
        \centering
        \includegraphics[width=\textwidth]{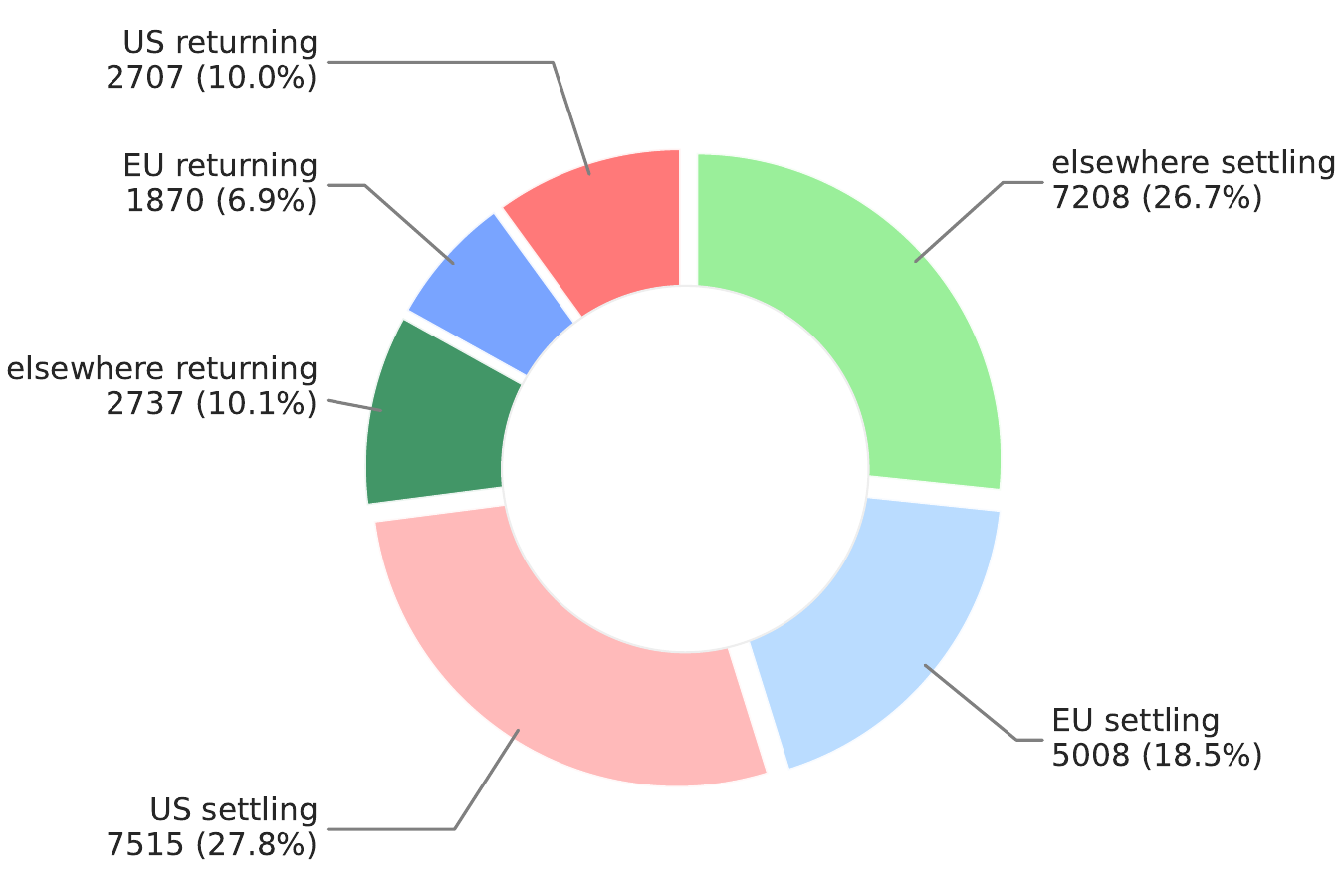}

        \label{fig:second_pie}
    \end{subfigure}

    \caption{\textbf{Frequency distribution of various categorized Indian researchers.}Authors in our dataset fall into three primary categories: researchers who never left India ($44.6\%$), researchers who began their research careers abroad ($38.2\%$), and researchers who started their careers in India but transitioned abroad at some point oftime ($17.2\%$) (left). Internationally mobile researchers are grouped into six categories based on their first destination to the US, EU, and elsewhere and whether they returned to India or settled there(right). Among these mobile researchers, the majority settle in the US (28\%), followed by those settling elsewhere (26.7\%) and in the EU (18.7\%). Returning scientists are less common, with less than 10\% returning from each region.}
    \label{fig_transitions}
\end{figure}

To identify and analyze different scientific career trajectories, we categorized authors into three groups based on their mobility patterns: (i) those who never left India (non-mobile), (ii) those who began their research careers abroad (abroad-origin), and (iii) those who moved abroad after beginning their research careers in India (India-to-abroad). The distribution of these categorized authors is shown on the left-hand side of Figure~\ref{fig_transitions}. Further, we excluded researchers of abroad-origin ($60,162$) from our analysis, as our focus is on outbound mobility trajectories of scientists who started their careers in India, thus yielding us a final dataset of $97,309$ authors.
To capture the different career pathways of the Indian researcher, we classified Internationally mobile researchers into the following six mobility categories:
(1) \textbf{US Returning} – Researchers who moved first to the US and subsequently returned to India;
(2) \textbf{EU Returning} – Researchers who moved first to the EU and later returned to India;
(3) \textbf{Elsewhere Returning} – Researchers who moved first to other high-income countries outside the US and EU, and later returned to India;
(4) \textbf{US Settling} – Researchers who moved to the US and never returned to India;
(5) \textbf{EU Settling} –  Researchers who moved to the EU and never returned to India;
(6)\textbf{Elsewhere Settling} – Researchers who moved first to other high-income countries outside the US and EU and did not return.
These mobility pathways are inferred from reconstructed affiliation histories. The distribution of mobile pathways is shown on the right-hand side of Figure~\ref{fig_transitions}. Considering both mobile and non-mobile trajectories, we find that $44.6\%$ of authors never move abroad, forming the core non-mobile control group in our study.  




\subsection*{Performance Evaluation of Career Trajectories using Matched Pair Analysis}

We assess each author’s career-age performance using three standardized metrics, which are the following (1) \textit{\textit{Citations per paper per career year}}, defined as the mean citations accrued per work up to that career age and it is calculated by dividing the total number of citations accrued up to a given career age by the total number of works published to that point. (2) \textit{productivity per career year}, the  number of works published per career year. It is computed by counting the number of works published in a career year. (3) \textit{Internationality}, the percentage of co-authors affiliated outside India at the publication time. For each paper, this is computed as the share of non-India-based co-authors, and is then aggregated across all papers authored by the individual to obtain an author level measure. On OpenAlex API, \textit{citation counts per year data} is available only from $2012$ onward, restricting this analysis to $47,779$ authors who began publishing in or after that year. Productivity is calculated for the full dataset of $97,308$ authors. \textit{Internationality} is computed for publications from $1980$ onward, covering $92,687$ authors. Career mobility trajectories, however, are reconstructed using the full OpenAlex dataset without restriction on starting year, ensuring that mobility patterns reflect the complete span of available data.
To determine the typical timing of international transitions, we compute the median age of first departure from India across all mobile trajectories and find it to be $7$ years since first publication. For matched-pair comparisons, each mobile trajectory is paired with immobile authors sharing the same career age and performance league (based on citations or productivity at that 7-year median). This matching strategy ensures that any observed post-transition divergence in performance can be attributed to mobility effects, rather than pre-existing differences.
For returnees, we also calculate the median age of return to India, excluding the settling trajectories (US Settling, EU Settling, and Elsewhere Settling),which by definition have no return transition.
\textit{Internationality} over time is analyzed using five trajectory-level indicators which are as follows:
(i) value at the transition age;
(ii) initial increase relative to the pre-transition minimum;
(iii) post-transition decline up to career age 40;
(iv) relative mean decrease in \textit{internationality} comparing returnees with non-returning peers who moved to the same region;
(v) relative mean retention of international co-authorship comparing returnees with immobile authors during the same post-transition window. 

\section*{Results}
Scientific mobility trajectories offer a data-driven perspective on key phenomena at both national and individual levels. At the national scale, they uncover patterns of brain drain, brain gain, and the contributions of researchers to Indian science. At the individual level, mobility shapes career progression through its impact on citations, productivity, and international collaboration. These insights reveal how international movement influences both personal careers and the broader scientific ecosystem.
In the following sections, we examine these mobility trajectories in detail and interpret the key findings from our statistical analysis of researcher career paths.
Please note that P-value significance levels are denoted as: $***$ for $p < 0.001$, $**$ for $p < 0.01$, and $*$ for $p < 0.05$ throughout the manuscript.
\subsection*{Mobility Patterns, Return Rates, Brain Drain  and Institutional Origins}
We visualize the distribution of career trajectories reconstructed for 1,57,471 Indian-affiliated researchers in Figure~\ref{fig_transitions}. The researchers who started abroad constituting $38.2\%$ of the total, have been excluded from the study, while almost half ($44.6\%$) of these researchers never publish from an institution outside India, and the remaining  $17.2\%$ undertake at least one international move during their careers. Among internationally mobile researchers, the US emerges as the most common destination (28\%), followed by non-EU countries (26.7\%) and the EU countries(18.7\%).
Return rate of of researcher to India is relatively rare than the departer rate from India. We find that only $26.6\%$ of internationally mobile researchers eventually return to India, reflecting a persistent and pronounced brain drain (Figure~\ref{fig_transitions}). In other words, a large majority of Indian-origin scientists who move abroad do not come back. The non-return or settling rates are consistently high across destinations, for example settling rate in the US is approximately $65\%$, $64\%$ for the EU, and $63\%$ for other countries.

Our analysis reveals that internationally mobile researchers predominantly originate from India's premier institutions, including top academic institutions like the Indian Institute of Science (IISc), the Indian Institutes of Technology (IITs), and the Indian Institutes of Science Education and Research (IISERs), as shown in Supplementary Information (SI) Figure S1. These findings underscore the need for policy reforms to support the scientific community, enabling India to retain and reintegrate the talent trained at its top institutions. Table S1 in SI presents the top 20 Indian institutes, ranked by the average number of international transitions per year, highlighting their role as key gateways to foreign institutions. However, the career trajectories of researchers are significantly dependent on the institution from which they transition, suggesting that institutional structures and academic environments influence long-term mobility outcomes. A matched-pair analysis illustrating these variations is provided in Figure S2 of the SI.


\subsection*{Effect on Citations}
\textit{Citations per paper per career year} serve as a key metric for evaluating the research impact of an individual. To investigate how international mobility influences this metric, we conducted a matched-pair analysis across different mobility categories. To isolate the effect of mobility from prior performance, we stratified authors based on their citation rates at the time of transition (see Methods). Specifically, researchers are grouped into three citation bands:[0–2], [2–3], and [3–5].
\textit{citations per paper per career year} at the time they left India. For each citation band, we calculate the average \textit{citations per paper per career year} for each author within a given trajectory, and then take the mean across all authors. 
Within each band, we compare transitioning authors with immobile peers matched for both citation range and career age. 

\begin{figure}[hbt]
\centering
            \includegraphics[width=1\linewidth]{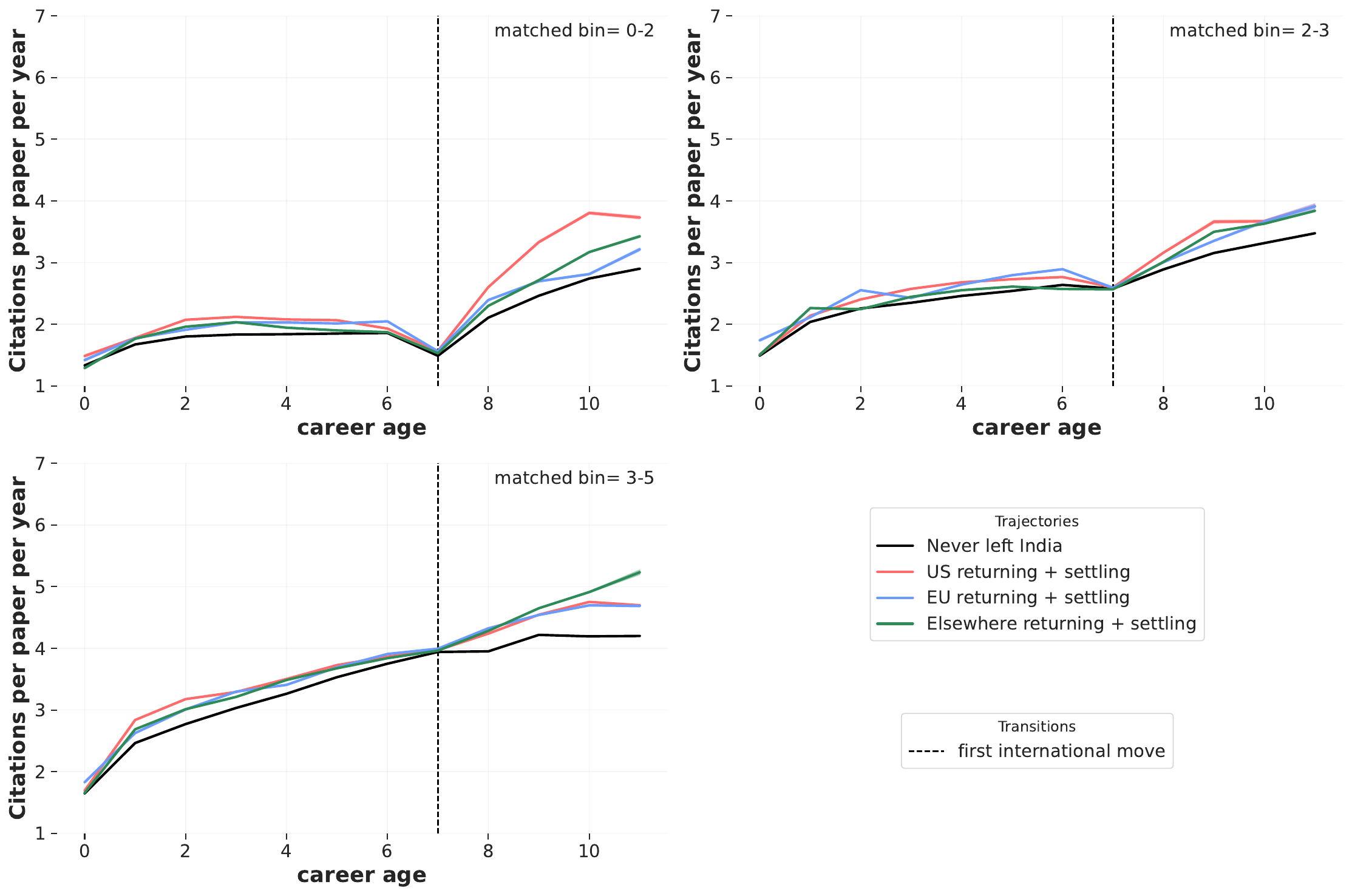}
            \caption{\textbf{Effect of mobility on scientific impact.} \textit{Citations per paper per career year} as a function of career age across mobility classes. The plot displays mean annual citations for scientists who remained in India and those who moved to the US, EU, or elsewhere, distinguishing between returning and settling trajectories. Shaded bands represent 95\% confidence intervals.The Vertical dashed line marks the median ages of first transition out of India (7 years since first publication). After this point, mobile researchers exhibit a steeper rise in citations per career year compared with immobile peers.}
            \label{fig:citations}
\end{figure}

Across all citation bands, Figure~\ref{fig:citations} shows that citation trajectories diverge immediately following international transition, with mobile authors accruing citations at a faster rate than their immobile counterparts. All mobile cohorts exhibit a clear and sustained increase in annual citation rates after the median departure age of seven years.
Because citation data are only available from 2012 onward, distinguishing returnees from settlers makes the sample sizes within each sub-category too small for any statistically valid comparisons. Thus, for citation analysis, we combine returning and settling authors into a single category. Two-sample Kolmogorov–Smirnov tests \cite{kolmogorov1933,smirnov1948} confirm that the citation distributions of all mobile groups differ significantly from those of their immobile counterparts (p < 0.001; see Table~\ref{table:citations1_updated},\ref{table:citations2_updated},\ref{table:citations3_updated}). These findings indicate that international mobility is associated with an immediate and sustained increase in research impact, likely facilitated by access to high-quality research environments and global collaboration networks. Statistical analyses across various research domains is presented in Tables S2 to S5 in the SI file, demonstrating the robustness of our findings when controlling for the topic of publication.

\begin{table}[hbt!]
\centering
\begin{tabular}{lllrr}
\hline
                         & \textbf{\small{Null Hypothesis Rejected}} & \textbf{\small{p-Value}} & \textbf{\small{Statistic}}   & \textbf{\small{Sample Size}} \\
\hline
US Returning         & True                     & *** & 0.1301080    & 1931        \\
EU Returning         & True                     & *** & 0.0890188   & 1611        \\
Elsewhere returning  & True                     & *** & 0.0937034   & 2364        \\
\hline
\end{tabular}
\caption{\textbf{KS 2-sample tests for citations (0–2 citations/year)} Kolmogorov–Smirnov results comparing mobile researchers to non-mobile peers within the 0–2 citations/year group.}
\label{table:citations1_updated}
\end{table}

\begin{table}[hbt!]
\centering
\begin{tabular}{lllrr}
\hline
                     & \textbf{\small{Null Hypothesis Rejected}} & \textbf{\small{p-Value}} & \textbf{\small{Statistic}}   & \textbf{\small{Sample Size}} \\
\hline
US Returning         & True                     & *** & 0.0821547   & 2386        \\
EU Returning         & True                     & *** & 0.0959079   & 1903        \\
Elsewhere returning  & True                     & *** & 0.0646384   & 2493        \\
\hline
\end{tabular}
\caption{\textbf{KS 2-sample tests for citations (2–3 citations/year)} Kolmogorov–Smirnov results comparing mobile researchers to non-mobile peers within the 2–3 citations/year group.}
\label{table:citations2_updated}
\end{table}

\begin{table}[hbt!]
\centering
\begin{tabular}{lllrr}
\hline
                     & \textbf{\small{Null Hypothesis Rejected}} & \textbf{\small{p-Value}} & \textbf{\small{Statistic}}   & \textbf{\small{Sample Size}} \\
\hline
US Returning         & True                     & *** & 0.0663874   & 4044        \\
EU Returning         & True                     & *** & 0.0744736   & 2875        \\
Elsewhere returning  & True                     & *** & 0.0741967   & 4392        \\
\hline
\end{tabular}
\caption{\textbf{KS 2-sample tests for citations (3–5 citations/year)} Kolmogorov–Smirnov results comparing mobile researchers to non-mobile peers within the 3–5 citations/year group.}
\label{table:citations3_updated}
\end{table}
\subsection*{Effect on Productivity}
\begin{figure}[t!]
    \centering
    \includegraphics[width=1\linewidth]{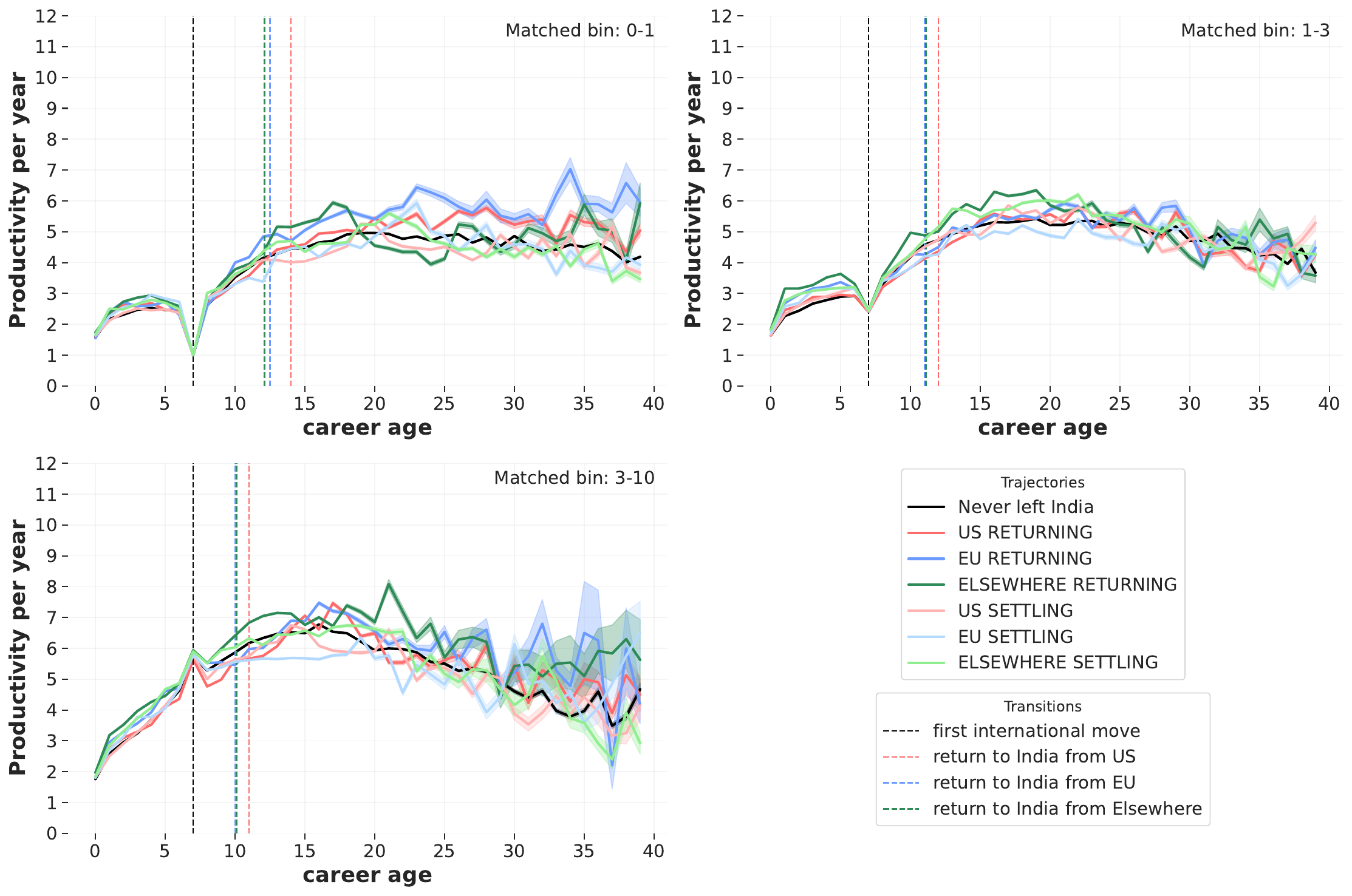}
    \caption{ \textbf{Effect of mobility on productivity.} The plot displays the number of articles published per career year per author 
    for each career age across all author groups-- Never left India (Immobile), US Returning, US Settling, EU Returning, EU Settling, Elsewhere Returning, and Elsewhere Settling. The shaded bands represent 95\% confidence intervals.Vertical dashed lines mark the median ages of first transition out of India (7 years since first publication) and the returns.}
    \label{fig:productivity}
\end{figure}

The productivity of researchers, measured as the \textit{number of publications per career year}, is used to assess whether international mobility leads to sustained increases in research output. Matched-pair analysis compares mobile researchers to non-mobile peers within the same productivity band and career age. In Fig.\ref{fig:productivity},  we average the number of papers published by an author at a career age averaged across all authors in that trajectory. Our results capture a short-term spike in productivity during the transition year for all mobility categories. However, unlike the clear citation gains observed after mobility, the long-term productivity differences between mobile and immobile researchers are not so clear. Two-sample Kolmogorov–Smirnov tests (Table \ref{table:productivity1},\ref{table:productivity2},\ref{table:productivity3}) show that researchers settle in the US publish on an average one to three papers per year, as well as those settling in the EU publish three to ten papers per year on an average, do not exhibit significant differences in productivity compared to non-mobile peers with the same output levels.
Interestingly, the most notable productivity gains are observed among low-output researchers (those publishing fewer than one paper per year) who transition to the US or EU. This suggests that international mobility may provide a short-term boost in productivity for early-stage or under performing researchers. For most other cohorts, productivity after transition largely realigns with that of non-mobile peers, indicating that international moves do not generally lead to sustained increases in research output over time.
All other mobile trajectories show statistically significant differences from their non-mobile counterparts. Although the p-values obtained in our statistical tests (Table~\ref{table:productivity1},\ref{table:productivity2},\ref{table:productivity3}) are statistically significant, the corresponding test statistics are quite small ($\leq 0.1$ in most cases, and $\leq 0.05$ for higher productivity cohorts). This indicates that, despite statistical significance, the magnitude of differences in productivity between mobile and non-mobile researchers is not strong.



\begin{table}[hbt!]
\centering
\begin{tabular}{lllrr}
\hline
                     & \textbf{\small{Null Hypothesis Rejected}}   & \textbf{\small{p-Value}}   &   \textbf{\small{Statistic}} &   \textbf{\small{Sample Size}} \\
\hline
 US Returning        & True                      & ***       &   0.0970788 &          6566 \\
 EU Returning        & True                      & ***       &   0.11107   &          4086 \\
 Elsewhere returning & True                      & ***       &   0.0774987 &          5035 \\
 US Settling         & True                      & ***       &   0.0246607 &         11597 \\
 EU Settling         & True                      & **        &   0.0273599 &          6713 \\
 Elsewhere Settling  & True                      & ***       &   0.0313334 &          9923 \\
\hline
\end{tabular}
\caption{\textbf{KS 2-sample tests for productivity (0–1 papers/year)} Kolmogorov–Smirnov results comparing mobile researchers to non-mobile peers within the 0–1 papers/year productivity group.}
\label{table:productivity1}
\end{table}

\begin{table}[hbt!]
\centering
\begin{tabular}{lllrr}
\hline
                     & \textbf{\small{Null Hypothesis Rejected}}   & \textbf{\small{p-value}}   & \textbf{\small{KS Statistic}} & \textbf{\small{Sample Size}} \\
\hline
US Returning         & True                       & ***       & 0.05         & 8158         \\
EU Returning         & True                       & ***       & 0.07         & 5829         \\
Elsewhere Returning  & True                       & ***       & 0.10         & 7624         \\
US Settling          & True                       & ***       & 0.02         & 14534        \\
EU Settling          & False                      & 0.52      & 0.01         & 9366         \\
Elsewhere Settling   & True                       & ***       & 0.03         & 12273        \\
\hline
\end{tabular}
\caption{\textbf{KS 2-sample tests for productivity (1–3 papers/year)} Kolmogorov–Smirnov results comparing mobile researchers to non-mobile peers within the 1–3 papers/year productivity group.}
\label{table:productivity2}
\end{table}

\begin{table}[hbt!]
\centering
\begin{tabular}{lllrr}
\hline
                     & \small{\textbf{Null Hypothesis Rejected}}   & \textbf{\small{p-value}}   & \textbf{\small{KS Statistic}} & \textbf{\small{Sample Size}} \\
\hline
US Returning         & True                       & **        & 0.03         & 6663         \\
EU Returning         & True                       & ***       & 0.04         & 4969         \\
Elsewhere Returning  & True                       & ***       & 0.05         & 7233         \\
US Settling          & True                       & ***       & 0.03         & 12143        \\
EU Settling          & True                       & ***       & 0.03         & 8403         \\
Elsewhere Settling   & False                      & 0.08      & 0.01         & 12653        \\
\hline
\end{tabular}
\caption{\textbf{KS 2-sample tests for productivity (3–10 papers/year)} Kolmogorov–Smirnov results comparing mobile researchers to non-mobile peers within the 3–10 papers/year productivity group.}
\label{table:productivity3}
\end{table}

\clearpage


\vspace{0.2cm}

\vspace{0.2cm}


\subsection*{Effect on Internationality}

\begin{figure}[ht!]
    \centering
    \includegraphics[width=1\linewidth, trim=0mm 0mm 0mm 1mm, clip]{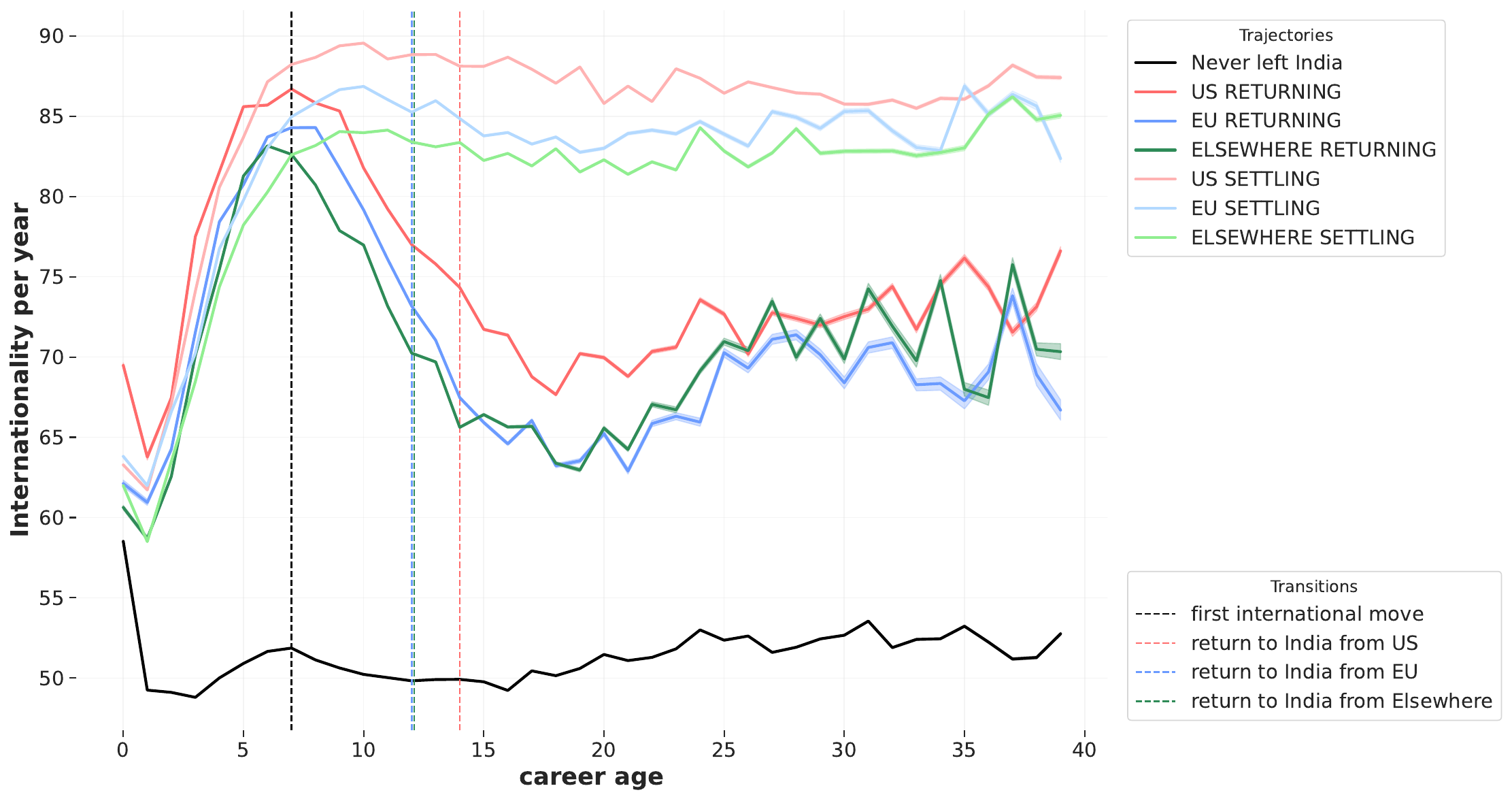}
    \caption{\textbf{Effect of mobility on internationality of collaboration networks of Indian researchers.} The curves in the plot shows the mean percentage of co-authors affiliated outside India (define as internationality) from career age $0$ to $40$ for seven trajectories: Immobile, US Returning, US Settling, EU Returning, EU Settling, Elsewhere Returning, and Elsewhere Settling. Color-coded curves ar corresponding to the various mobility trajectory as mentioned in the labels and shaded bands indicating $95\%$ confidence intervals. Mobile researchers exhibit a sharp rise in internationality at transition, with settling groups maintaining higher levels thereafter, while returning groups peak and then decline but remain above the Immobile baseline.}
    \label{fig:internationality}
\end{figure}

Internationality is measured as the \textit{proportion of an Indian researcher's co-authors affiliated with institutions outside India at the time of article publication}. This metric captures the networking benefits associated with international mobility. Figure~\ref{fig:internationality} illustrates variations in internationality across mobility categories, averaged across all authors at that career age for the specific trajectory. To systematically assess internationality changes with mobility patterns, Table~\ref{table:internationality-comparison} reports four key metrics: internationality at the transition point, the immediate increase relative to pre-transition values, the post-return decline, and the long-term retention rate compared to the immobile control group. Researchers who move abroad exhibit a sharp rise in international co-authorship, with internationality increasing by at least 26.5\% at the median age of transition. The average share of non-Indian collaborators rises from a baseline of 52\% (among immobile researchers) to over 80\% across all returnee categories. Among six mobile categories, the three settling categories sustain their elevated levels of international collaboration throughout their careers. In contrast, the three returning categories experience a post-return decline in internationality up to $20.44\%$ after returning to India. Despite this decline and a subsequent plateau around career age $20$, these researchers continue to maintain significantly higher international collaboration rates ($32$ to $35$ percentage points higher) than peers who never left India. Kolmogorov–Smirnov two-sample tests (Table~\ref{table:internationality}) confirm that the distributions of internationality for all mobile groups are statistically distinct ($p < 0.001$) from the immobile baseline. These tests, which are computed beyond the median transition age, allow us to robustly reject the null hypothesis that mobile and immobile trajectories follow similar collaboration patterns. Field-specific analyses (Tables S2–S5 in the SI file) further support the generality of these trends across disciplines. In conclusion, this analysis underscores that international mobility plays a critical role in enhancing global research collaboration, with its effects often persisting throughout the duration of a researcher's career.

\vspace{0.25cm}
\begin{table}[ht]
\centering
\begin{tabular}{lrrrr}
\hline
\small{\textbf{Metric}} & \small{\textbf{Never left India}} & \small{\textbf{US returning}} & \small{\textbf{EU returning}} & \small{\textbf{Elsewhere returning}} \\
\hline
Value at transition age (\%)       & 52.00 & 87.00 & 84.00 & 83.00 \\
Initial increase (\%)              & 5.76  & 26.50 & 27.72 & 29.01 \\
Post-transition decrease (\%)      & 0.67  & 16.62 & 19.34 & 16.36 \\
Relative mean decrease (\%)        & --    & 20.40 & 23.92 & 20.44 \\
Relative mean retention (\%)       & --    & 40.32 & 32.02 & 33.92 \\
\hline
\end{tabular}
\caption{International collaboration metrics for returnees versus non-mobile researchers at comparable career ages.}
\label{table:internationality-comparison}
\end{table}

\vspace{0.25cm}

\begin{table}[ht]
\centering
\begin{tabular}{l c c c r}
\hline
\small{\textbf{Category}}               & \small{\textbf{Null hypothesis rejected}} & \small{\textbf{p-value}} & \small{\textbf{KS statistic}} & \small{\textbf{Sample size}} \\
\hline
US Returning           & Yes                      & ***      & 0.39         & 19,993      \\
EU Returning           & Yes                      & ***      & 0.33         & 13,437      \\
Elsewhere Returning    & Yes                      & ***      & 0.33         & 18,313      \\
US Settling            & Yes                      & ***      & 0.58         & 45,251      \\
EU Settling            & Yes                      & ***      & 0.52         & 27,212      \\
Elsewhere Settling     & Yes                      & ***      & 0.50         & 39,561      \\
\hline
\end{tabular}
\caption{\textbf{KS 2-sample tests for internationality} Kolmogorov–Smirnov test results comparing international collaboration rates of each mobility category against researchers who never left India.}

\label{table:internationality}
\end{table}

\subsection*{Robustness of Mobility Effects Across Disciplinary Boundaries}
To examine whether various research domains influences the observed trends, we replicate our analysis across $12$ distinct research areas defined in the OpenAlex repository. Topic-specific results for fields such as Computer Science, Medicine, and Engineering are presented in Figures S3-S6 of SI. These analyses confirm the robustness of our core findings across disciplinary boundaries. While some variation exists, certain fields exhibit slightly higher or lower levels of citation impact and international collaboration across all mobility categories, however overall differences are small. Importantly, the general patterns observed in the aggregated data persist within individual disciplines, indicating that our conclusions are not driven by the topical composition of the sample.

\section*{Discussion}

This study presents a system-level analysis of how international mobility shapes the career trajectories of researchers affiliated with India. Drawing on affiliation histories from the OpenAlex database, we focus on profiles of authors associated with Indian institutions, categorizing them broadly into immobile and internationally mobile groups. Our analysis reveals that only 27\% of internationally mobile researchers eventually return to India, indicating a dominant pattern of one-way mobility. This sustained trend of brain drain underscores the urgent need for effective domestic strategies to retain scientific talent. Policy interventions aimed at strengthening research infrastructure, expanding funding opportunities, and enhancing institutional support could play a pivotal role in reversing this trend and encouraging return migration. Notably, we find that the median age at first international transition is seven years after a researcher’s first publication, typically corresponding to the postdoctoral or early faculty stage a critical juncture for career development.

 Further we link International mobility to three core dimensions of scientific performance: citation impact, publication productivity, international collaboration. We find that international mobility is associated with a persistent increase in citation impact. Citations per paper rise sharply at the moment of departure and remain approximately $35\%$ higher than those of non-mobile peers for at least a decade. This citation advantage suggests that early access to global research environments significantly increases visibility and amplifies scientific influence.
 Mobility is also linked to a short-term increase in publication output during the year of transition, but this effect is not sustained. Post-mobility productivity levels quickly converge back to those of non-mobile counterparts. Analysis of the impact of mobility on the international collaboration networks of Indian researchers reveals that those who return to India after working abroad retain significantly higher levels of international collaboration. On average, returnees maintain a 32–40 percentage point advantage in foreign co-authorship compared to their non-mobile peers, underscoring their role as critical bridges connecting Indian science to global research networks. These patterns hold consistently across 12 major research domains defined in the OpenAlex database, highlighting the robustness of this finding irrespective of scientific discipline.

Overall, our study presents one of the most comprehensive, data-driven analyses of scientific mobility patterns among Indian-affiliated researchers. We uncover the long-term effects of international mobility on citation impact, productivity, and internationality. Our findings offer valuable insights into the dynamics of brain drain, return migration, and international engagement, with important implications for national science policy and institutional planning. In particular, the study highlights how mobility boosts citation impact and international collaboration, especially among returnees, underscoring the critical role these researchers play in linking Indian science to global networks. These results can inform strategies for talent retention and repatriation in India and other developing countries facing similar challenges.

Our study has some limitations that warrant consideration. First, citation data in the OpenAlex API are only available from $2012$ onward, potentially biasing impact estimates toward more recent careers and underrepresenting earlier cohorts. Second, affiliation histories are inferred algorithmically by OpenAlex and without manual-checks, some short-term visits or overlapping institutional ties may still be misclassified. Third, we restrict our analysis to authors with $10$ – $200$ publications to ensure data stability, which may exclude early-career researchers and hyper-prolific individuals, thereby affecting the generalizability of our estimates. Additionally, the dataset lacks demographic variables such as gender, caste, or institutional prestige, limiting equity-based assessments. Finally, our collaboration metric is based solely on co-authorship, omitting other forms of collaboration such as informal mentorship, grant participation, industry partnerships etc..

In the future, we can focus to enhance the resolution of mobility studies by integrating curricula vitae, institutional records, grant data, and online public profiles to improve disambiguation and demographic coverage. Our findings also suggest for comparative cross-national analyses. For instance, do other emerging economies exhibit similar patterns of one-way mobility toward the US and EU and other first-world economy countries? Comparative research could explore gendered mobility trajectories, the interplay between funding access and migration, and whether returnees catalyze new domestic scientific networks. The inclusion of additional metrics, such as altmetrics, patent activity, and institutional rankings would allow for a broader assessment of how mobility affects both scholarly impact and innovation capacity. Ultimately, combining large-scale bibliometric data with AI-driven entity resolution and enrichment methods could enable the creation of a global observatory of scientific careers. Such an infrastructure would allow policymakers and institutions to move beyond anecdotal accounts and toward evidence-based strategies for fostering brain circulation rather than enduring brain drain.

\section*{Data availability}
The data from \href{openalex.org}{openalex.org} used in this work is openly accessible for download using the API \href{https://api.openalex.org/works}{https://api.openalex.org/works}.

\section*{Code availability}
The code used in this study is available at
\href{https://github.com/SirajTM2199/Researcher-Networks-Plotting}{https://github.com/SirajTM2199/Researcher-Networks-Plotting}.

\bibliography{refs}

\section*{Acknowledgments}
CM acknowledge financial support from DST, India (INSPIRE Grant No. IFA19-PH248) and IISER Thiruvananthapuram for providing computing facilities. 


\section*{Additional information}

\begin{center}
{\LARGE \textbf{Supplementary Information }}\\[0.7cm]
\end{center}

\subsection*{Prestigious Indian institutes}
India’s research ecosystem is anchored by several key national institutions, each with distinct missions and areas of emphasis. Many of these institutes are widely recognized for offering integrated academic and research programs, with student admissions based on competitive national-level entrance examinations. For example, the Indian Institute of Science (IISc) and the Indian Institutes of Science Education and Research (IISERs/NISERs) primarily focus on basic science research and integrated science education. In contrast, the Indian Institutes of Technology (IITs) emphasize engineering, applied sciences, and technology-driven research. These institutions play a critical role in shaping the country’s scientific landscape by fostering research excellence and training the next generation of scientists and engineers.

In addition to these national institutes, several state and central universities, along with specialized research organizations such as the Academy of Scientific and Innovative Research (AcSIR) and laboratories under the Council of Scientific and Industrial Research (CSIR), have also established strong reputations for high-quality research programs. Collectively, these institutions contribute significantly to India’s overall scientific output and innovation capacity, often serving as hubs of domain-specific expertise and international collaboration.

To assess how institutional affiliation influences research career trajectories, we grouped researchers based on their entry-point institutions within the Indian academic system. We define four cohorts:
(i) researchers who began their careers at the Indian Institutes of Technology (IITs), which are premier and old institutions for technical education and applied research;
(ii) those initially affiliated with the Indian Institute of Science (IISc), India’s oldest and most prestigious center for advanced scientific research, known for its strong interdisciplinary focus;
(iii) researchers from the IISER/NISER system, which was designed to integrate high-quality science education with foundational research training in the basic sciences and they are newer compared to IISc and some of the IITs; and
(iv) researchers affiliated with institutions outside the IIT/IISER/NISER/IISc framework, which, despite receiving less national visibility, have demonstrated considerable research productivity and impact.

Beyond the institutions included in our cohort analysis, it is important to acknowledge other key players that contribute uniquely to India's research output. Notable among these include 

the Vellore Institute of Technology (VIT), representing the growing influence of high-performing private universities in technology and applied sciences; and All India Institutes of Medical Sciences, Delhi (AIIMS), the nation's foremost network for medical education and biomedical research. While these institutions are significant contributors to the scientific landscape, a starting institute analysis was not performed on them for this particular study due to their distinct organizational structures and specialized mandates.
\begin{figure}[ht]
    \centering
    \includegraphics[width=1\linewidth]{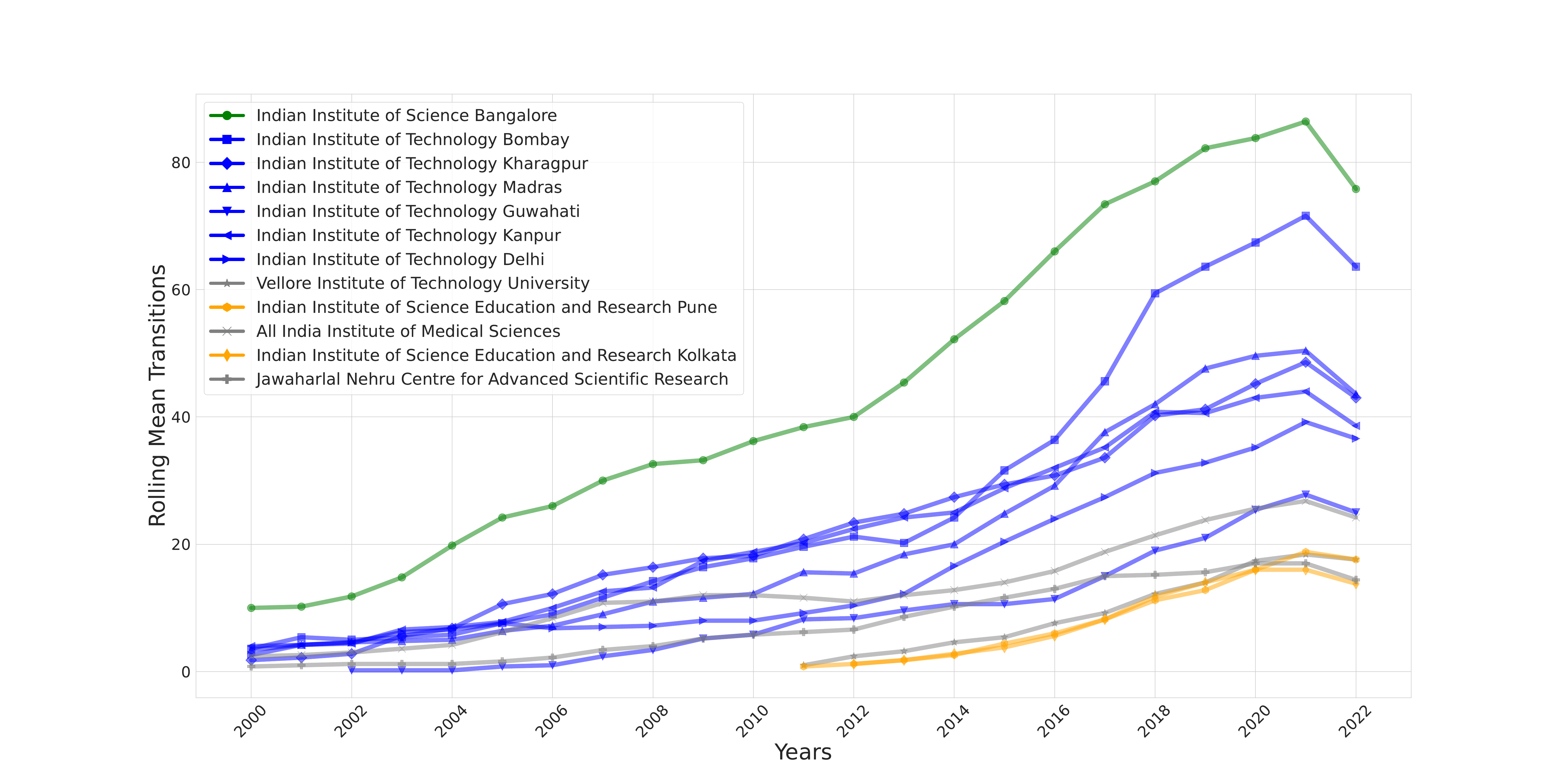}
    \caption{Top 12 Indian Institutes with the most transitions abroad. The green colored line represents IISc, the blue ones are all IITs, oranges are the IISERs; and the three gray lines are respectively from top to bottom All India Institute of Medical Sciences, Delhi, Vellore Institute of Technology, Chennai and Jawaharlal Nehru Center for Advanced Scientific Research, Bangalore. }
    \label{Figure:Institute_level_transitions}
\end{figure}

\begin{figure}
    \centering
    \includegraphics[width=1.0\linewidth]{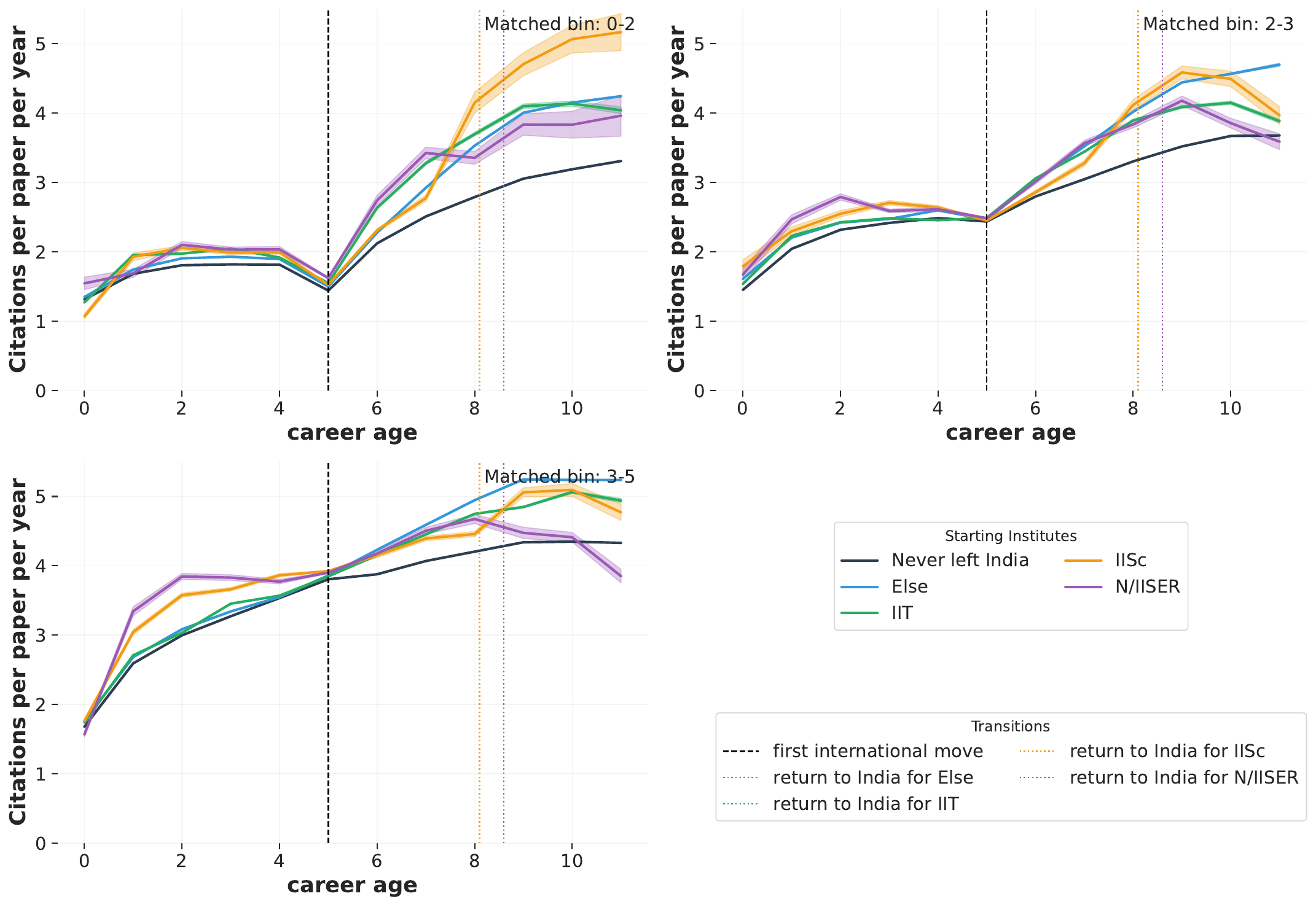}
    \caption{Citations per paper vs Career age, annual citations for different starting institutes. The orange line represents the citation trajectory for researchers who started out in IISc, with year 0 marking the point of their first move abroad. The purple line shows the corresponding trajectory for the researchers who started out in IISERs and NISERs, the green line for the researchers who began their research careers from IITs, and the black line represents immobile authors as the control group, and finally, the blue line represents researchers whose careers began in other institutions within India. Shaded bands represent $95\%$ confidence intervals.}
    \label{figure:MPA_Institutions}
\end{figure}

\subsection*{Measuring foreign Mobility from top $20$ Indian Institutes}
In our analysis, we examine the top 20 Indian institutions and universities based on our the dataset. While many of these belong to the well-established IIT/IISER/NISER/IISc framework, several high-performing institutions outside these categories also emerge as significant contributors to India’s scientific landscape. These institutions play a vital role in advancing the nation’s research output and in training future generations of scientists—often matching, and in some cases exceeding, the performance of their more widely recognized counterparts.

To examine the institutional drivers of international scientific mobility from India, we analyze the flow of researchers from Indian institutions to institutions abroad. Figure S\ref{Figure:Institute_level_transitions} illustrates the number of outbound transitions per year from the top 12 Indian universities and research institutes, based on their representation in our dataset. 
\tref{tab:top_institutions_abbr} highlights the total count, Transitions per year and the US/EU ratio for each institute from an expanded list of the top 20 Institutes in India from our study. The ranking has been made on the basis of Transitions per year, US/EU ratio refers to the number of transitions to the US for each transition to the EU, a notable observation is that research institutes such as IISERs and NISERs show lower US/EU ratios, while institutions such as IITs show greater transitions to the US.
These transitions reflect the movement of researchers who begin their careers at Indian institutions and later take positions at foreign institutions, capturing patterns of cross-border academic mobility. This analysis highlights the key contributors to India’s scientific diaspora and reveals which Indian institutions serve as major launchpads for international careers. By focusing on annual trends, the figure also sheds light on the temporal dynamics of outbound mobility and offers insights into how different institutions contribute to global knowledge exchange over time.

\begin{table}[htbp]
\centering
\caption{Top 20 Indian Institutions by Research Metrics}
\label{tab:top_institutions_abbr}
\begin{tabular}{rlrrr}
\toprule
\small{\textbf{Rank}} & \small{\textbf{Institution Name}} & \small{\textbf{Count}} & \small{\textbf{Transitions / Year}} & \small{\textbf{US/EU Ratio}} \\
\midrule
1 & IISc Bangalore & 1227 & 26.67 & 1.49 \\
2 & IIT Bombay & 750 & 22.73 & 1.65 \\
3 & IIT Kharagpur & 608 & 16.89 & 1.51 \\
4 & IIT Madras & 570 & 15.41 & 1.35 \\
5 & IIT Guwahati & 255 & 13.42 & 1.50 \\
6 & IIT Kanpur & 606 & 12.89 & 1.44 \\
7 & IIT Delhi & 452 & 12.56 & 1.32 \\
8 & VIT, Chennai & 146 & 10.43 & 1.31 \\
9 & IISER Pune & 135 & 10.38 & 0.73 \\
10 & AIIMS & 360 & 10.00 & 3.50 \\
11 & IISER Kolkata & 116 & 8.92 & 0.83 \\
12 & JNCASR, Bangalore & 204 & 8.16 & 1.10 \\
13 & TIFR, Mumbai & 355 & 8.07 & 1.12 \\
14 & IIT Roorkee & 250 & 8.06 & 1.67 \\
15 & IISER Mohali & 80 & 8.00 & 1.00 \\
16 & IISER Bhopal & 86 & 7.82 & 0.86 \\
17 & Jadavpur University, Kolkata & 248 & 7.75 & 2.02 \\
18 & JNU, Delhi & 235 & 7.58 & 2.60 \\
19 & IIT Hyderabad & 105 & 7.50 & 1.78 \\
20 & IACS, Kolkata & 296 & 7.40 & 1.35 \\
\bottomrule
\vspace{0.3mm}
\end{tabular}
\parbox{\linewidth}{\small \textit{Note:} IISc: Indian Institute of Science; IIT: Indian Institute of Technology; VIT: Vellore Institute of Technology; IISER: Indian Institute of Science Education and Research; AIIMS: All India Institute of Medical Sciences; JNCASR: Jawaharlal Nehru Centre for Advanced Scientific Research; TIFR: Tata Institute of Fundamental Research; JNU: Jawaharlal Nehru University; IACS: Indian Association for the Cultivation of Science.}
\end{table}

Among the many research and academic institutions in India, the premier institutes include the IISERs, NISER, IITs, and IISc. Our analysis reveals distinct career-stage patterns in international mobility across these institutional categories. As shown in Figure S\ref{figure:MPA_Institutions} the joint category of IISERs and NISER dominates during the early stages of a researcher’s career, reflecting their undergraduate research-oriented curriculum and strong emphasis on academic publishing at the student level. In contrast, mobility associated with IISc increases later in a researcher’s career, consistent with its role as a leading destination for doctoral and postdoctoral training. The IITs and other major institutions exhibit relatively flatter curves, suggesting more uniform patterns of mobility over the career span. 

This variation aligns with the differing institutional missions: while IISERs and NISER focus on early academic training in the basic sciences, the IITs have a stronger orientation toward engineering and applied sciences, often with industry-facing outputs. As a result, researchers from the IITs and similar institutions may be less likely to have substantial publication records early in their careers compared to those from research-intensive basic science institutions.

Our findings further show that institutions of national importance play a dominant role in India’s outbound scientific mobility. In particular, the Indian Institute of Science (IISc) emerges as the leading platform for international transitions, followed closely by the IITs. Figure S~\ref{Figure:Institute_level_transitions} illustrates these trends by depicting the average annual number of transitions from selected Indian institutions to institutions in high-income countries since the year 2000. Each curve represents a specific institution and captures its yearly contribution to outbound mobility. IISc consistently outpaces other institutions, with IIT Bombay also showing a marked increase in international placements in recent years.

\section*{Mobility-Driven Impact Across Fields}
\begin{figure}
    \centering
    \includegraphics[width=1\linewidth,trim=0mm 0mm 0mm 0mm]{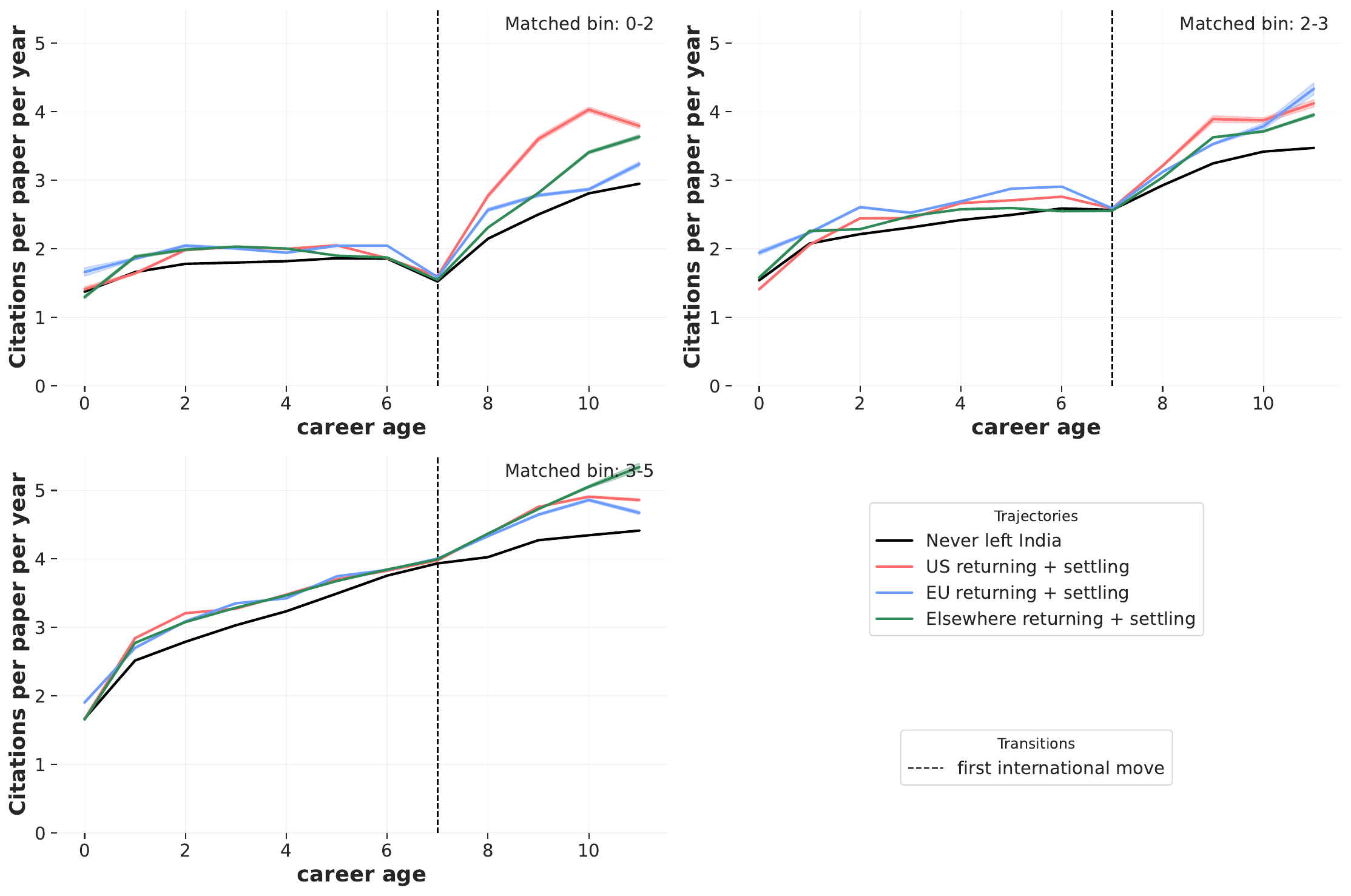}
    \caption{Citations per paper by career age across mobility categories in Computer Science. Dashed line indicates median age of international transition (7 years). Shaded regions are 95\% confidence intervals. The matched bins are different bands of researchers based on the citations per paper per career year they had the time of transitioning abroad.}
    \label{fig:cs_citations}
\end{figure}

\begin{figure}
    \centering
    \includegraphics[width=1\linewidth]{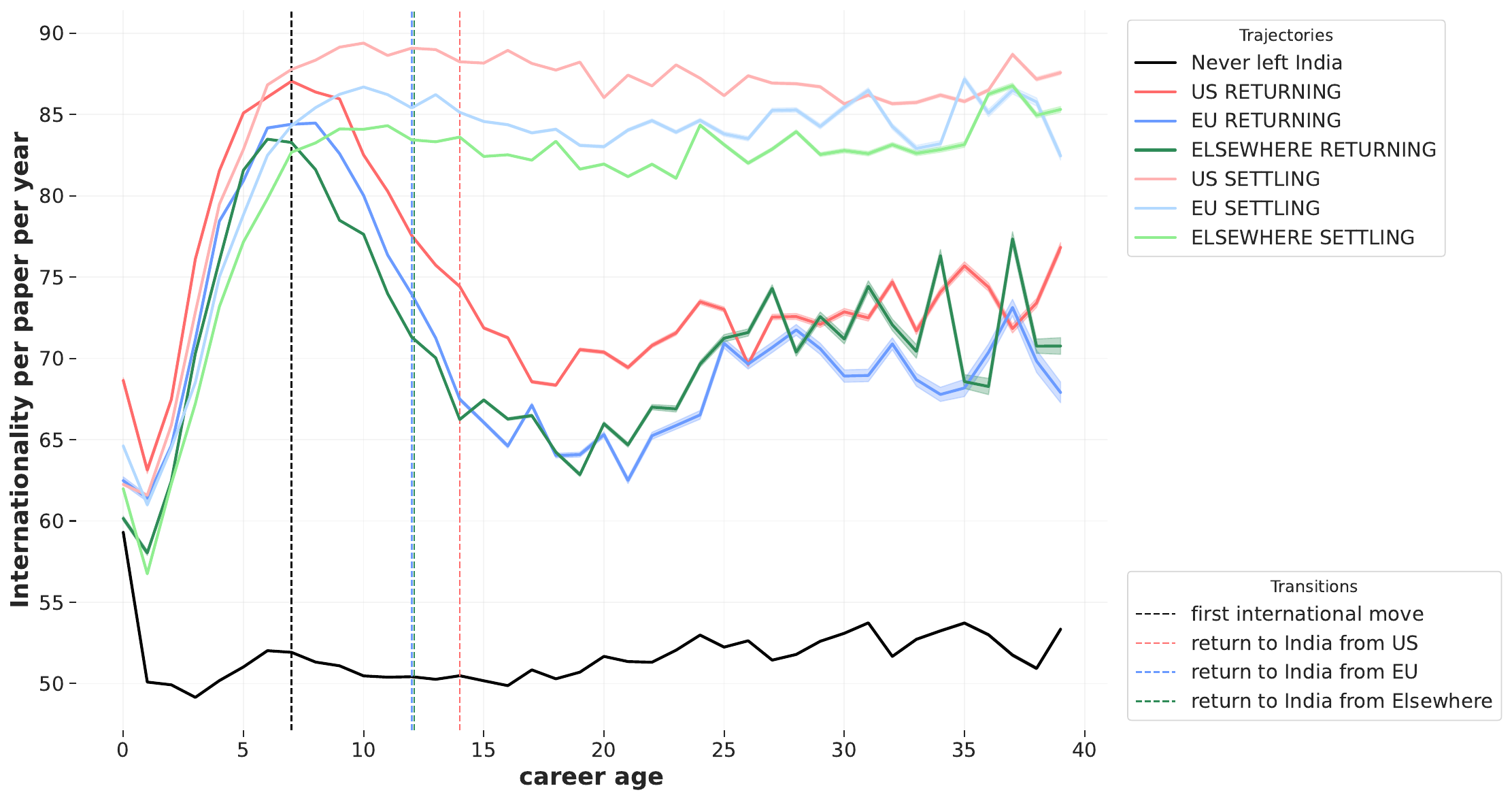}
    \caption{International co-authorship share by career age across mobility categories in Computer Science.}
    \label{fig:cs_internationality}
\end{figure}

\begin{figure}
    \centering
    \includegraphics[width=1\linewidth]{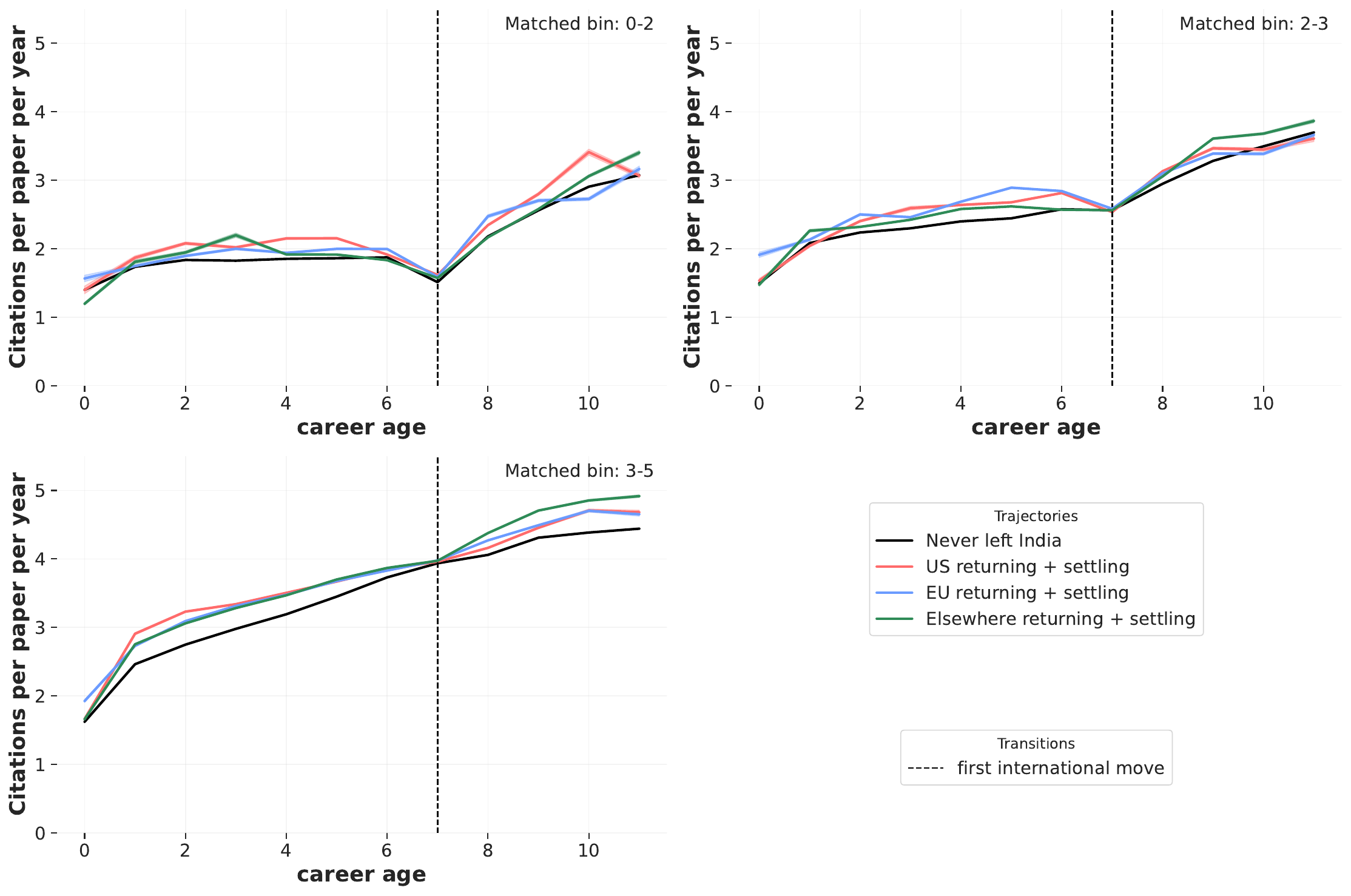}
    \caption{Citations per paper by career age across mobility categories in Medicine.}
    \label{fig:med_citations}
\end{figure}

\begin{figure}
    \centering
    \includegraphics[width=1\linewidth]{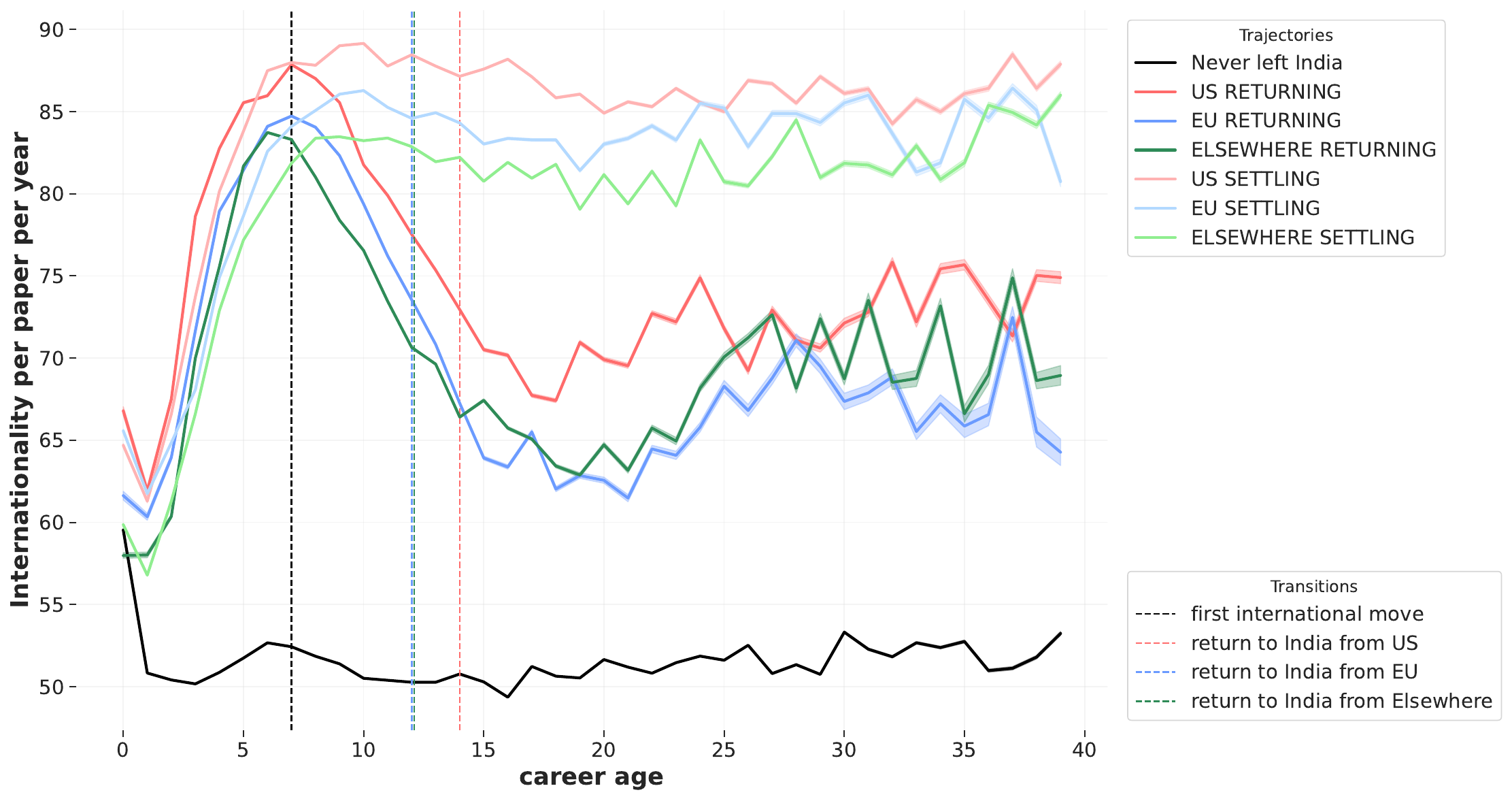}
    \caption{International co-authorship share by career age across mobility categories in Medicine).}
    \label{fig:med_internationality}
\end{figure}

To evaluate whether the effects of scientific mobility are consistent across research disciplines, we conducted a focused comparison between two distinct subject areas: Computer Science and Medicine. These fields were selected for their contrasting publication cultures and collaborative norms, offering a broader perspective on how mobility shapes research careers across scientific domains. Figures S~\ref{fig:cs_citations} and S~\ref{fig:med_citations} show the citation trajectories, while Figures S~\ref{fig:cs_internationality} and S~\ref{fig:med_internationality} present international collaboration patterns for these two domains.

Our analysis reveals that the structure of scientific career trajectories is more strongly influenced by the mobility category i.e., whether a researcher remained in India, settled abroad, or returned than by the specific research domain. While some field-level variation exists, particularly in absolute citation counts or baseline internationality, the overall trends remain consistent: mobile researchers tend to experience higher citation impact and maintain broader international collaboration networks compared to their non-mobile peers, regardless of disciplinary affiliation.

This finding is reinforced by our broader analysis across twelve major subjects in the OpenAlex classification: Physics, Chemistry, Biology, Engineering, Computer Science, Medicine, Mathematics, Environmental Science, Social Sciences, Materials Science, Business/Management and Health Professions. In each domain, researchers in the 'settled abroad' and 'returnee' categories exhibit enhanced citation and internationality metrics compared to those who remained in India. These results confirm that mobility, rather than field of research, is the dominant factor shaping long-term research impact and global engagement.

\clearpage
\section*{Statistical Analysis of Scientific Trajectories: Topic-wise Assessment}

\tref{table:citations_0_2} through \tref{table:internationality_sup} present the results of the statistical analysis done by performing the Kolmogorov-Smirnov tests on the values of citations per paper per career age and internationality after the median age of transition ($7$ years after the first publication)evaluating the influence of two key factors, citation impact and international collaboration on the likelihood of Indian researchers returning from abroad. To enhance clarity, the data are segmented into separate tables based on citation rate bands and levels of internationality. For each subgroup, the tables report whether the null hypothesis is rejected, along with the corresponding p-value and test statistic. These analyses allow us to assess the statistical significance of differences between groups.

Specifically, we compare the career trajectories of researchers who returned to India after working in the United States or the European Union with those who remained in India throughout their careers. By limiting the comparisons to US-returning and EU-returning cohorts, we focus on the two most common destinations for Indian researchers abroad, enabling a clearer interpretation of return dynamics in relation to impact and collaboration indicators.  

The results indicate that the in all citation bands, the null hypothesis has been rejected at conventional significance levels (see note below) confirming that the distributions of returnees, settlers and non-mobile researchers differ significantly in terms of career age wise citations per paper and fraction of international collaborations. Thus, statistically giving further weight to our findings that transitioning abroad has demonstrable benefits to the career of a researcher. 

\noindent \small
\textbf{Note:} P-value significance levels are denoted as: $***$ for $p < 0.001$, $**$ for $p < 0.01$, and $*$ for $p < 0.05$.


\begin{longtable}{@{}l l l r r@{}}
\caption{Statistical Analysis for researchers with 0-2 Citations} \label{table:citations_0_2} \\
\toprule
\small{\textbf{Topic}} & \small{\textbf{Trajectory}} & \small{\textbf{Null Rejected}} & \small{\textbf{P-Value}} & \small{\textbf{Statistic}} \\
\midrule
\endfirsthead

\multicolumn{5}{c}{Table S\ref{table:citations_0_2} -- continued from previous page} \\
\toprule
\textbf{Topic} & \textbf{Trajectory} & \textbf{Null Rejected} & \textbf{P-Value} & \textbf{Statistic} \\
\midrule
\endhead

\bottomrule
\multicolumn{5}{r@{}}{{Continued on next page}} \\
\endfoot

\bottomrule
\endlastfoot

Medicine & US Returning & True & *** & 0.102351 \\
& EU Returning & True & * & 0.0627125 \\ \midrule
Social Sciences & US Returning & True & *** & 0.102776 \\
& EU Returning & True & *** & 0.0879958 \\ \midrule
Engineering & US Returning & True & *** & 0.118566 \\
& EU Returning & True & *** & 0.0962582 \\ \midrule
Arts and Humanities & US Returning & True & *** & 0.110234 \\
& EU Returning & True & *** & 0.0835126 \\ \midrule
Computer Science & US Returning & True & *** & 0.133674 \\
& EU Returning & True & *** & 0.104657 \\ \midrule
Biochemistry, Genetics & US Returning & True & *** & 0.106878 \\
& EU Returning & True & *** & 0.0711101 \\ \midrule
Agricultural/Biological Sci. & US Returning & True & *** & 0.109721 \\
& EU Returning & True & *** & 0.0732777 \\ \midrule
Environmental Science & US Returning & True & *** & 0.108536 \\
& EU Returning & True & *** & 0.0816076 \\ \midrule
Physics and Astronomy & US Returning & True & *** & 0.118006 \\
& EU Returning & True & * & 0.0617174 \\ \midrule
Materials Science & US Returning & True & *** & 0.111637 \\
& EU Returning & True & *** & 0.0702207 \\ \midrule
Business/Management & US Returning & True & *** & 0.11373 \\
& EU Returning & True & *** & 0.0743939 \\ \midrule
Health Professions & US Returning & True & *** & 0.107783 \\
& EU Returning & True & *** & 0.0767698 \\ 

\end{longtable}

\clearpage

\begin{longtable}{@{}l l l r r@{}}
\caption{Statistical Analysis for Academics with 2-3 Citations} \label{table:citations_2_3} \\
\toprule
\small{\textbf{Topic}} & \small{\textbf{Trajectory}} & \small{\textbf{Null Rejected}} & \small{\textbf{P-Value}} & \small{\textbf{Statistic}} \\
\midrule
\endfirsthead
\multicolumn{5}{c}{\tablename\ \thetable{} -- continued from previous page} \\
\toprule
\textbf{Topic} & \textbf{Trajectory} & \textbf{Null Rejected} & \textbf{P-Value} & \textbf{Statistic} \\
\midrule
\endhead
\bottomrule
\multicolumn{5}{r@{}}{{Continued on next page}} \\
\endfoot
\bottomrule
\endlastfoot

Medicine & US Returning & True & *** & 0.0697922 \\
& EU Returning & True & *** & 0.0833347 \\ \midrule
Social Sciences & US Returning & True & *** & 0.0833474 \\
& EU Returning & True & *** & 0.0949636 \\ \midrule
Engineering & US Returning & True & *** & 0.0911911 \\
& EU Returning & True & *** & 0.1051 \\ \midrule
Arts and Humanities & US Returning & True & *** & 0.0817051 \\
& EU Returning & True & *** & 0.0960403 \\ \midrule
Computer Science & US Returning & True & *** & 0.0908266 \\
& EU Returning & True & *** & 0.124684 \\ \midrule
Biochemistry, Genetics & US Returning & True & *** & 0.0743892 \\
& EU Returning & True & *** & 0.0986754 \\ \midrule
Agricultural/Biological Sci. & US Returning & True & *** & 0.0795056 \\
& EU Returning & True & *** & 0.102748 \\ \midrule
Environmental Science & US Returning & True & *** & 0.0796246 \\
& EU Returning & True & *** & 0.0912785 \\ \midrule
Physics and Astronomy & US Returning & True & *** & 0.0888989 \\
& EU Returning & True & *** & 0.0976311 \\ \midrule
Materials Science & US Returning & True & *** & 0.090145 \\
& EU Returning & True & *** & 0.103872 \\ \midrule
Business/Management & US Returning & True & *** & 0.0842107 \\
& EU Returning & True & *** & 0.0953711 \\ \midrule
Health Professions & US Returning & True & *** & 0.0818499 \\
& EU Returning & True & *** & 0.100291 \\ 
\end{longtable}

\clearpage

\begin{longtable}{@{}l l l r r@{}}
\caption{Statistical Analysis for Academics with 3-5 Citations} \label{table:citations_3_5} \\
\toprule
\small{\textbf{Topic}} & \small{\textbf{Trajectory}} & \small{\textbf{Null Rejected}} & \small{\textbf{P-Value}} & \small{\textbf{Statistic}} \\
\midrule
\endfirsthead
\multicolumn{5}{c}{\tablename\ \thetable{} -- continued from previous page} \\
\toprule
\textbf{Topic} & \textbf{Trajectory} & \textbf{Null Rejected} & \textbf{P-Value} & \textbf{Statistic} \\
\midrule
\endhead
\bottomrule
\multicolumn{5}{r@{}}{{Continued on next page}} \\
\endfoot
\bottomrule
\endlastfoot

Medicine & US Returning & False & 0.02 & 0.0358745 \\
& EU Returning & True & *** & 0.0583472 \\ \midrule
Social Sciences & US Returning & True & *** & 0.0629126 \\
& EU Returning & True & *** & 0.0667087 \\ \midrule
Engineering & US Returning & True & *** & 0.0728135 \\
& EU Returning & True & *** & 0.0570824 \\ \midrule
Arts and Humanities & US Returning & True & *** & 0.0627895 \\
& EU Returning & True & *** & 0.0673648 \\ \midrule
Computer Science & US Returning & True & *** & 0.0647458 \\
& EU Returning & True & *** & 0.0675199 \\ \midrule
Biochemistry, Genetics & US Returning & True & ** & 0.0478448 \\
& EU Returning & True & *** & 0.0594126 \\ \midrule
Agricultural/Biological Sci. & US Returning & True & *** & 0.0623925 \\
& EU Returning & True & *** & 0.0683278 \\ \midrule
Environmental Science & US Returning & True & *** & 0.0648001 \\
& EU Returning & True & *** & 0.0640534 \\ \midrule
Physics and Astronomy & US Returning & True & *** & 0.0626306 \\
& EU Returning & True & *** & 0.0778263 \\ \midrule
Materials Science & US Returning & True & *** & 0.0585445 \\
& EU Returning & True & *** & 0.0765215 \\ \midrule
Business/Management & US Returning & True & *** & 0.0632462 \\
& EU Returning & True & *** & 0.0663162 \\ \midrule
Health Professions & US Returning & True & *** & 0.0635495 \\
& EU Returning & True & *** & 0.0683431 \\ 
\end{longtable}

\clearpage

\begin{longtable}{@{}l l l r r@{}}
\caption{Statistical Analysis for Internationality Factor} \label{table:internationality_sup} \\
\toprule
\textbf{Topic} & \textbf{Trajectory} & \textbf{Null Rejected} & \textbf{P-Value} & \textbf{Statistic} \\
\midrule
\endfirsthead
\multicolumn{5}{c}{\tablename\ \thetable{} -- continued from previous page} \\
\toprule
\small{\textbf{Topic}} & \small{\textbf{Trajectory}} & \small{\textbf{Null Rejected}} & \small{\textbf{P-Value}} & \small{\textbf{Statistic}} \\
\midrule
\endhead
\bottomrule
\multicolumn{5}{r@{}}{{Continued on next page}} \\
\endfoot
\bottomrule
\endlastfoot

Medicine & US Returning & True & *** & 0.393289 \\
& EU Returning & True & *** & 0.334441 \\ \midrule
Social Sciences & US Returning & True & *** & 0.395418 \\
& EU Returning & True & *** & 0.337125 \\ \midrule
Engineering & US Returning & True & *** & 0.384342 \\
& EU Returning & True & *** & 0.321914 \\ \midrule
Arts and Humanities & US Returning & True & *** & 0.391223 \\
& EU Returning & True & *** & 0.332557 \\ \midrule
Computer Science & US Returning & True & *** & 0.385345 \\
& EU Returning & True & *** & 0.331305 \\ \midrule
Biochemistry, Genetics & US Returning & True & *** & 0.381902 \\
& EU Returning & True & *** & 0.3273 \\ \midrule
Agricultural/Biological Sci. & US Returning & True & *** & 0.390299 \\
& EU Returning & True & *** & 0.330545 \\ \midrule
Environmental Science & US Returning & True & *** & 0.392811 \\
& EU Returning & True & *** & 0.334792 \\ \midrule
Physics and Astronomy & US Returning & True & *** & 0.395071 \\
& EU Returning & True & *** & 0.328605 \\ \midrule
Materials Science & US Returning & True & *** & 0.391632 \\
& EU Returning & True & *** & 0.334578 \\ \midrule
Business/Management & US Returning & True & *** & 0.393308 \\
& EU Returning & True & *** & 0.335375 \\ \midrule
Health Professions & US Returning & True & *** & 0.391528 \\
& EU Returning & True & *** & 0.332729 
\end{longtable}


\end{document}